# Association between Prefrontal fNIRS signals during Cognitive tasks and College scholastic ability test (CSAT) scores: Analysis using a quantum annealing approach

Yeaju Kim[1,†], Junggu Choi[1,†], Bora Kim[2], Yongwan Park[3], Jihyun Cha[4], Jongkwan Choi[4], and Sanghoon Han[1,5,*]

[1]Yonsei Graduate Program in Cognitive Science, Yonsei University, Seoul 03722, Republic of Korea

[2]Department of Counselling, Honam University, Gwangju 62399, Republic of Korea

[3]Department of Business Administration, Gyeongsang National University, Jinju 52828, Republic of Korea

[4]OBELAB Inc., Seoul 06211, Republic of Korea

[5]Department of Psychology, Yonsei University, Seoul 03722, Republic of Korea

[*]Corresponding author: Sanghoon Han (sanghoon.han@yonsei.ac.kr)

[†]Yeaju Kim and Junggu Choi contributed equally to this study.

## Abstract

**Background:** Academic achievement is a critical measure of intellectual ability, prompting extensive research into cognitive tasks as potential predictors. Neuroimaging technologies, such as functional near-infrared spectroscopy (fNIRS), offer insights into brain hemodynamics, allowing understanding of the link between cognitive performance and academic achievement. Herein, we explored the association between cognitive tasks and academic achievement by analyzing prefrontal fNIRS signals.

**Methods:** A novel quantum annealer (QA) feature selection algorithm was applied to fNIRS data to identify cognitive tasks correlated with CSAT scores. Twelve features (signal mean, median, variance, peak, number of peaks, sum of peaks, slope, minimum, kurtosis, skewness, standard deviation, and root mean square) were extracted from fNIRS signals at two time windows (10- and 60-second) to compare results from various feature variable conditions. The feature selection results from the QA-based and XGBoost regressor algorithms were compared to validate the former's performance. In a three-step validation process using multiple linear regression models, correlation coefficients between the feature variables and the CSAT scores, model fitness (adjusted $R^2$), and model prediction error (RMSE) values were calculated.

**Results:** The quantum annealer demonstrated comparable performance to classical machine learning models, and specific cognitive tasks, including verbal fluency, recognition, and the Corsi block tapping task, were correlated with academic achievement. Group analyses revealed stronger associations between Tower of London and N-back tasks with higher CSAT scores.

**Conclusions:** Quantum annealing algorithms have significant potential in feature selection using fNIRS data, and represents a novel research approach. Future studies should explore predictors of academic achievement and cognitive ability.

**Keywords**

Academic achievement assessment, Functional near-infrared spectroscopy, Cognitive tasks, Quantum annealer, Machine learning

## Introduction

Since the first intelligence quotient (IQ) test was developed in 1905, numerous scholars and organizations have contributed to the objective measurement and assessment of academic achievement.[1] Under the broader theme of evaluating human intellectual ability, researchers have investigated several potential predictors of academic achievement, such as intelligence and cognitive abilities. While classic approaches mainly focus on general intelligence and its positive correlation with academic achievement levels,[2,3] recent methods have attempted to identify and predict academic achievement based on the concept of multidimensional cognitive ability.[4,5] Accordingly, several cognitive tasks have been proposed for such an evaluation. The N-back task and Tower of London (ToL) task are both examples of cognitive tasks that assess advanced cognitive abilities such as working memory capacity or executive planning.[6,7] Previous studies[8,9] have explored the relationship between cognitive task performance and academic achievement.

When assessing the results of cognitive tasks, basic dependent measure generally comprises behavioral performance, such as error rate or reaction time. In addition, many researchers[10,11] have studied brain hemodynamics among participants. Neuroimaging technologies, such as functional magnetic resonance imaging (fMRI) and functional near-infrared spectroscopy (fNIRS), allow the quantification of neural activation and connectivity patterns during specific cognitive tasks. In one prior study, Herff et al.[12] examined mental workload activation levels in the prefrontal cortex using fNIRS signals during the N-back task phase. fNIRS measures cortical activation using near-infrared rays reflected from changes in oxy- and deoxyhemoglobin concentrations, with significant advantages in terms of higher temporal resolution, cost efficiency, and portability.[13] Previous studies[14,15] have investigated the relationship between cognitive performance and brain hemodynamics in the prefrontal regions, as well as their implications for academic achievement examinations.

The analysis of highly multidimensional neural dynamics data, statistical models, and machine learning models have been widely used because of their specialization in multivariate pattern analysis.[16,17] In one such study,[18] researchers identified a possible biomarker of pain using feature selection through machine learning in fNIRS dataset. Noori et al.[19] selected optimal feature sets from fNIRS signals based on a genetic algorithm that is widely applied to feature selection. Huang et al.[20] proposed connectivity-based feature selection methods for stress detection in fNIRS signals. Aydin[21] discussed the importance of feature selection for achieving improved algorithm performance. Recently, a quantum annealer-based feature selection algorithm was introduced and applied to neuroimaging data analysis.[22,23] Quantum annealer (QA) is a novel optimization method that specializes in combinatorial optimization executed using a quantum computer. Optimized combinations are sought in the feature set by minimizing a cost function, i.e., finding the ground state of the energy in a quantum system to find the global minimum in optimization tasks.[24] Although this was not the first attempt to employ a quantum-annealer-based method for neuroimaging data analysis, research investigating the relative importance of cognitive tasks using fNIRS data remains scarce.

In this study, we investigated the relationship between cognitive tasks and academic achievement using brain hemodynamics data measured with fNIRS devices. One hundred college students performed seven different cognitive tasks (with a preceding resting session) and fNIRS signals were collected in each task phase. Twelve features (signal mean, median, variance, peak, number of peaks, sum of peaks, slope, minimum, kurtosis, skewness, standard deviation, and root mean square) were extracted from the fNIRS signals collected for each task. Among them, features that were highly correlated with the College scholastic ability test (CSAT) scores were selected and ranked using QA feature selection algorithms. To further validate the results from the quantum annealer, we compared their performances with those of classical machine learning (CML) models (i.e., support vector machine, decision tree, and XGBoost regressor). The XGBoost regressor was selected for further analysis because it exhibited the best performance among the three regressors. The process for the performance validation of the QA feature selection method comprised three steps using a multiple linear regression (MLR) model, as follows: (1) correlation between features and the dependent variable (i.e., CSAT score), (2) model fitness of the MLR model, and (3) model prediction error in the MLR model. Finally, the cognitive tasks most highly associated with academic performance were selected based on the results of the validation process.

# Methods

## Overview of the study

The following six steps were conducted to investigate the relationship between academic achievement and brain hemodynamics during cognitive task performance: First, fNIRS signals were collected from 100 college students during the performance of eight different cognitive tasks (including a resting-state session). Second, the fNIRS dataset was preprocessed to remove the artifacts and noise of the dataset. Third, 12 different features (signal mean, median, variance, peak, number of peaks, sum of peaks, slope, minimum, kurtosis, skewness, standard deviation, and root mean square) were extracted from the preprocessed oxy-hemoglobin (HbO) and deoxy-hemoglobin (HbR) concentration signals during each of the eight experimental tasks. Fourth, the QA-based algorithm and XGBoost regressor utilized the 12 extracted feature sets to investigate the features that were highly associated with the average CSAT score. Fifth, we validated the ranked features from the QA and XGBoost feature selection algorithms based on the correlation coefficient between the features and the dependent variable, MLR model fitness, and MLR model prediction error. Finally, the cognitive tasks that were highly correlated with the CSAT scores were selected based on the results of the validation process. An overview of this study is presented in Figure 1.

## Experimental procedure

The experimental procedure during data collection was as follows: After the participants were provided with a brief introduction to the fNIRS signal collection procedure and provided written consent, the researchers informed them that the experiment comprised eight independent tasks, with a short break after every three tasks. During fNIRS data acquisition, the participants were instructed not to move their heads as much as possible.

First, before performing the seven cognitive tasks, participants underwent a 5-minute resting session without any instructed cognitive tasks while wearing an fNIRS device. Following a resting session, seven cognitive tasks (i.e., the Corsi block-tapping task, emotion task, recognition task, Stroop task, Tower of London task, N-back task, and verbal fluency task) were presented sequentially. Second, the Corsi block-tapping (CBT) task was performed to measure short-term visuospatial working memory. In this task, participants were asked to indicate consecutive locations of colored squares.[25] Third, participants were instructed to indicate the emotional state of the presented facial stimuli. Six choices (disgust, sadness, happiness, anger, surprise, or fear) were provided for each facial expression. Fourth, a recognition task was conducted to investigate whether participants correctly recognized the former faces and their emotional states. The pictures from the previous task and a set of new face pictures were presented in a randomized order, and participants were instructed to respond to 1) whether they recognized the face and, if yes, 2) which emotional state the faces had beforehand. Fifth, the participants performed the Stroop task, which is widely used to measure overall executive functions, such as selective attention and response inhibition.[26] Participants were presented with colored words and instructed to choose the correct color/meaning of the given words. Sixth, the Tower of London (ToL) task was performed. The ToL task, in which participants change the location of balls in a limited number of attempts, is often applied to measure executive planning and problem-solving abilities.[7] Seventh, an N-back test was used to measure working memory capacity.[27] In this task, the participants were asked to memorize the sequential position of a target stimulus and respond if its current position matched the position of the previous n-trials (hence, n-back). Lastly, the verbal fluency task (VFT) was performed to measure participants' verbal memory and cognitive control.[28] Participants were asked to produce as many words as they could verbally under three different conditions (control, phonemic, and semantic).

The participants were compensated with 20,000 Korean Won per hour. All experiments were conducted following approval of the Institutional Review Board of Yonsei University (7001988–202104-HR-659-06).

## fNIRS measurement and Preprocessing

One hundred undergraduate students participated in eight experimental tasks during fNIRS data collection (60 females, 40 males; mean age = 19.17 years). The students were recruited from 13 universities across South Korea. The participant pool was purposely distributed to diverse sources to broaden the range of collected cognitive performance levels. Data from seven participants were eliminated due to errors during the collection process, leaving 93 participants in total.

For data collection, we measured the prefrontal hemodynamic activity of the participants using the NIRSIT Lite device (OBELAB Inc., Seoul, Korea). This device comprises five dual-wavelength laser diodes (780 and 850 nm) and seven photodetectors, forming 15 channels with a laser-detector distance of 3 cm. The optical signal from each channel was sampled at 8.138 Hz. Figure 2 shows the positions of the optodes in the NIRSIT Lite device.[29]

After acquisition, raw fNIRS signals were preprocessed to eliminate physiological noise and channels with a low signal-to-noise ratio. The dataset was filtered using a bandpass range of 0.005–0.1 Hz. Channels with signal-to-noise ratios of < 30 dB were excluded from the final dataset. The relative changes in HbO and HbR concentrations for each task were obtained using the modified Beer-Lambert law.[30]

## Feature extraction

Feature extraction from the preprocessed HbO and HbR signal datasets was performed as described below. Because the duration of the eight cognitive tasks varied, we needed to trim the fNIRS data to control the number and temporal sizes of the windows. The task duration of the 'emotion' task (i.e., 211 seconds) was the shortest among 8 tasks, and thus we trimmed the signal data to remain only the first 211 seconds in all eight tasks. Each feature (i.e., signal mean, median, variance, standard deviation, root mean square, peak, minimum, slope, number of peaks, sum of peaks, skewness, and kurtosis) of each task phase was extracted from non-overlapping time windows of 10 s and 60 s. Figure 3 depicts how we extracted the 10-second time windows from the overall task duration. More detailed definitions and calculation formulas of the twelve features are as below:

*Signal Mean*

The mean values of the HbO and HbR signals in each time window were calculated using the following formula:

$$\mu_w = \frac{1}{N_w}\sum_{i=i_1}^{i_2} HbX(i) \quad (1)$$

To calculate $\mu_w$, the mean value for a given window, the summation of HbO and HbR signal data (denoted *HbX*) from the start point to the end point of the window was divided by the number of signal values in the window (denoted $N_w$). Mean signal values have been utilized in numerous fNIRS-BCI studies. For example, Hong et al.[31] used the mean signal features to classify brain hemodynamic data from individuals with cognitive and motor impairments.

*Signal Median*

The median values of HbO and HbR concentration signals were selected for each time window. Shin and Jeong[32] used the median features of fNIRS data signals from the motor cortex to classify the neural hemodynamic responses during different motor tasks.

*Signal Variance and Standard deviation*

The signal variance values of HbO and HbR signal were calculated as follows:

$$Variance_w = \frac{1}{N_w}\sum_{i=i_1}^{i_2} (HbX(i) - \mu_w)^2 \quad (2)$$

Where $Variance_w$ indicates the variance of feature values in each time window. Standard deviations were determined by taking the root of variance values. Multiple studies have reported the effectiveness of these features in fNIRS analysis, where neural data are classified into different task phases.[33,34]

*Signal Root Mean Square*

The signal root mean square values of the HbO and HbR signals in each time window were calculated as follows:

$$RMS_w = \sqrt{\frac{1}{N}\sum_{i = i_1}^{i_2} HbX(i)^2} \qquad (3)$$

The RMS feature values have previously been used in fNIRS analyses of emotion induction and language learning processes.[35,36]

*Signal Peak and Minimum*

Peak (highest) and minimum (lowest) values were extracted for each time window. Signal peak and minimum values have been used and validated in several fNIRS-BCI studies, mostly to improve the quality of BCI systems using classification methods.[37,38,39]

*Signal Slope*

The signal slopes were calculated as follows:[40]

$$Slope_w = H_w - L_w \qquad (4)$$

The differences between the highest (denoted $H_w$) and lowest point (denoted $L_w$) of the concentration signal values were calculated for each time window.

*Number and Sum of Signal Peaks*

The number of values and the sum of all signal peak values in each given window were calculated. Khan and Hong[41] calculated the number of peaks by measuring the local maxima of concentration changes in the HbO signals in each time window. The local maxima are summed in a given window to calculate the sum of the peak values.

*Signal Skewness and Kurtosis*

The skewness value $Skewness_w$ was calculated from the signal data in the given windows as follows:

$$Skewness_w = \frac{E_x(HbX_w - \mu_w)^3}{\sigma^3} \tag{5}$$

In this formula, the standard deviation (denoted $\sigma^3$) divides the expectation of HbO or HbR signal (denoted $E_X$). The signal kurtosis values (denoted $Kurtosis_w$) were calculated from the signal data in the given windows as follows:

$$Kurtosis_w = \frac{E_x(HbX_w - \mu_w)^4}{\sigma^4} \tag{6}$$

Several previous studies have used signal skewness and kurtosis values in their fNIRS studies, both in research on sensory input processing and mental imagery interpretation.[42,43]

## QA based feature selection method

To identify the features most strongly associated with the aforementioned feature sets with CSAT levels, a QA-based feature selection method was applied. This method is executed on quantum processing units (QPU) of D-Wave quantum computing hardware. A D-wave QPU was fundamentally built on the Ising model from statistical mechanics. The Ising model comprises nodes and edges in a lattice structure. In D-wave quantum hardware, each node was physically converted into a superconducting flux qubit, while two states of qubits (i.e., spin-up and spin-down) were converted into binary values (+1 and -1) in Ising models, respectively. The correlation values between the qubits were calculated to describe their relationships. The qubit states and relationships can be reflected in the objective function of the Ising model as follows:

$$f_{Ising}(s) = \sum_{i=1}^{N} h_i s_i + \sum_{i=1}^{N} \sum_{j=i+1}^{N} J_{i,j} s_i s_j \tag{7}$$

Where $h_i$ indicates the bias of the $i^{th}$ qubit, and $s_i$ represents the spin of the qubits in the D-wave QPU. The coupling strength between two qubits ($s_i$ and $s_j$) is denoted by $J_{i,j}$. This model aims to determine the ground state of a quantum system by minimizing the energy state through an optimization process.

Furthermore, the Ising model can be translated into binary quadratic models for diverse optimization problems. The objective function formula of BQM is similar to that of the Ising model.

$$f(x) = \sum_{i} Q_{i,i}x_i + \sum_{i<j} Q_{i,j}x_i x_j = x^T Q x \tag{8}$$

where $Q_{i,j}$ is elements in the $i^{th}$ row and $j^{th}$ column of a upper triangular matrix $Q$ ($Q \in M_{m\times n}(R)$) and $x_i$ represents $i^{th}$ binary values ($x_i \in \{0,1\}$) in $x$. We can determine the optimal solutions with a given coefficient matrix $Q$ by minimizing the objective function $f(x)$.

In this study, we calculated mutual information (MI) to reflect the correlation between features, and conditional mutual information (CMI) to apply the relationships between features and dependent variables. MI (8) and CMI (9) were defined as follows:

$$I(X;Y) = H(X) - H(X|Y) \tag{9}$$

$$I(X;Y|Z) = H(X|Z) - H(X|Y,Z) \tag{10}$$

where $H(X)$ denotes the marginal entropy of $X$. It quantifies the amount of information contained in features. And $H(X|Y)$ denotes the conditional entropy. Negative CMI values were used as diagonal terms, and MI values were applied as upper-triangular terms of the coefficient matrix $Q$.

The BQM library, a part of the D-Wave Ocean software, transforms the objective function into a graph $G = (V, E)$, where $V$ signifies a set of vertices and $E$ represents a set of edges connecting vertices encoding MIs and CMIs. This graph is embedded into the QPU of the D-Wave hardware. To determine the optimal feature combination, the BQM conducted an iterative optimization process for a specified number of features, denoted by $k$, ranging from 1 to 168 using 10-second windows (1 to 24 using 60-second windows). Upon completion of the entire process, the relative importance of the features was established by counting their occurrence. For a more comprehensive understanding of graph-level implementations in quantum hardware, please refer to Harris et al.[44].

## CML models

To examine the feature selection performance of the QA-based algorithms, we utilized CML models as counterparts. Three CML algorithms (support vector machine, XGBoost regressor, and decision tree regressor) were compared to determine the relevant classical counterpart. The support vector machine (SVM) regressor was the first classification algorithm employed. This supervised learning method specializes in predicting continuous outcomes by employing kernel functions to transform data into high-dimensional spaces. This transformation facilitates the discovery of regression functions that correspond to nonlinear relationships within the dataset. The primary aim is to maximize the margin between data points and a regression line, defining an epsilon (ε)-insensitive tube to penalize deviations beyond this margin. Support vectors, which are crucial data points that shape the position of the regression line, are integral to this process. Optimization occurs by identifying the hyperplane that minimizes the errors within this specified margin.

The decision tree regressor, which is the second classification algorithm, operates by recursively partitioning the dataset into smaller subsets and creating a tree-like structure in which each internal node represents a feature, each branch signifies a decision based on that feature, and each leaf node provides a regression output. By selecting the most informative features at each step, typically based on criteria such as minimizing the variance or mean squared error, the algorithm constructs a tree to predict continuous target variables. The third algorithm was the XGBoost regressor. This algorithm is an ensemble-learning technique that sequentially builds decision trees to refine and enhance predictions. By combining multiple shallow decision trees, it creates a robust predictor adept at capturing complex relationships in datasets. Based on operational processes similar to the decision tree algorithms, the predicted values and errors were calculated for each branch of the tree structure.

Among the three CML algorithms, we selected the XGBoost regressor as a comparison method for the QA-based feature selection based on the regression performance of each algorithm. Detailed experimental results for the three CML models are presented in the Results section. From the two main feature selection methods (i.e., the QA-based feature selection method and XGBoost regressor for comparison), we needed to validate the feature selection results of the two methods with additional criteria.

## Validation process with MLR models

As described above, we ranked the feature variables (168 variables from 10 s windows; 24 variables from 60 s windows) collected during each task phase using QA-based feature selection and the CML model (i.e., XGBoost regressor) algorithms. Using these data, we examined the validity of each rank structure and compared the results of the two algorithms. The following three-step validation process was adopted to confirm that the results show the relative importance of each feature:

First, the correlation coefficient values between each feature from both algorithms and the average CSAT score were calculated. The correlation coefficient values were averaged for every ten feature variables, starting with the top 10. In other words, we calculated the mean correlation coefficient values for every ten feature variables listed on the rank in order (i.e., top 1-10, top 11-20, top 21-30, and so on). Highly ranked feature variables were expected to show higher correlation coefficients with average CSAT scores. Thus, this process was conducted to confirm that the coefficient values showed an increasing trend for highly ranked features.

In the following steps, MLR models were used to evaluate the model fitness and performance of each algorithm. Multiple linear regression (MLR) is a common statistical method used to investigate the linear relationship between a dependent variable and multiple independent variables.[45] To validate the proposed algorithms, the MLR models used the same features and dependent variables as those of the QA-based feature selection algorithms. With 168 variables from 10 s windows, five different variable conditions were set based on the accumulative rank of each variable (i.e., top 10, top 20, top 30, top 40, and top 50). The same excluding two-variable conditions (i.e., top 10 and top 20) were set for 24 variables from the 60 s windows.

Subsequently, the MLR model fitness of the two algorithms was compared to examine how accurately the ranked features predicted the dependent variables. We calculated the adjusted R-squared value of the MLR models to compare the model fitness of each feature selection algorithm, with and without the stepwise feature selection method. Higher model fitness values indicated that the model accurately captured the relatedness of the dependent variable (i.e., the CSAT score) and independent variables (i.e., selected features).

Finally, the root mean square error (RMSE) values were calculated to compare the regression performance (i.e., model prediction error) of the two algorithms. Lower RMSE values indicate that the models show better regression performance, which means that the feature ranks reflect the relative importance of the features.

### Tools

The fNIRS dataset was pre-processed using the NIRIST Lite Analysis Tool (version 3.3.0) and Python software (version 3.11; scikit-learn, version 1.2.1). We used D-Wave Ocean software (Python-based framework) to build a QA-based feature selection algorithm. The MLR models and XGBoost regressors were built and operationalized using Python (version 3.11; scikit-learn, version 1.2.1) and R (version 4.3.1).

## Results

### Algorithm performances of classical machine learning models

We calculated the evaluation metrics (RMSE values) of each of three CML regression models (the support vector machine, decision tree, and XGBoost regressor) to compare their performance. This process was conducted to select the model with the best performance for validating the performance of the QA algorithm. Twelve features of the pre-processed HbO and HbR signal data were used as independent variables to predict the average CSAT score for each participant in the regression process.

For all 12 features (signal mean, median, variance, peak, number of peaks, sum of peaks, slope, minimum, kurtosis, skewness, standard deviation, and root mean square), the XGBoost regressor showed the best performance (i.e., lowest RMSE values) among the three models for all twelve features. The average RMSE values of the models were as follows: support vector machine, 1.517; decision tree, 1.786; and XGBoost regressor, 1.229. Therefore, the feature selection results from the XGBoost regressor were employed in the validation process of the QA-based algorithm. The RMSE values of the three models are listed in Table 1.

## Selected features from the QA-based feature selection algorithm

The QA-based feature selection algorithm ranked 168 variables from 10 s windows and 24 variables from 60 s windows based on relatedness with the average CSAT score. The top ten variables in each feature window were considered when selecting the final cognitive task. The VFT, CBT, and recognition tasks were the most selected cognitive tasks in the top 10 rank order. Tables 2 and 3 show the top 10 results of the cognitive task variables for the 10 s and 60 s windows.

In Table 2, each cell contains information regarding the cognitive task that each feature variable represents, the time window the feature was extracted from (0-20), and the number of times each feature was selected during the feature selection process. For example, 'recog_9 (159)' indicates the feature from the tenth window of recognition task phase, which was selected 159 times by quantum annealing algorithm. Table 3 has the same format but the time windows range from 0-2.

## Selected features from the XGBoost regression algorithms

The same process was performed using the XGBoost feature selection algorithm. Again, 168 variables from the 10 s windows and 24 variables from the 60 s windows were ranked based on their relatedness to the dependent variable. Among the top ten variables in each ranked feature list, the VFT, recognition, and emotion tasks were the most frequently selected. Tables 4 and 5 show the top 10 results of the cognitive task variables for the 10 s and 60 s windows. In Table 4, each cell contains information regarding the cognitive task the feature variable represents, the time window the feature was extracted from (0-20), and the F-score values. Table 5 again has the same format but the time windows range from 0-2.

## Results of three steps in the validation process

In the validation process, we first calculated the correlation coefficient values of the extracted feature variables and average CSAT scores of each participant. As expected, the coefficient values decreased when lower-ranked features were added to the variables. Figures 4-7 show the overall trend of the selected features and average CSAT scores for both the QA-based and XGBoost algorithms.

Second, the model fitness values were calculated and compared between the MLR models based on each algorithm (QA and XGBoost). When the model fitness values for the five different variable conditions (i.e., top 10, 20, 30, 40, and 50) were compared, the adjusted $R^2$ values increased when more features were added to the models. In other words, there was an overall pattern of higher model fitness values in the 'top 50' condition, which used 50 feature variables to build the model, than in the 'top 10' condition, which used only 10 feature variables. Furthermore, when comparing the results from the QA algorithms with those of the XGBoost algorithm, we found that the features from the 60-second windows show higher similarity between the two algorithms. Tables 6 and 7 show the average model fitness (adjusted $R^2$) values for each model and time window.

In Table 6, each cell shows the fitness (adjusted $R^2$) values of the QA and XGBoost models under different variable conditions. The adjusted $R^2$ values calculated without the stepwise feature selection method are shown in parentheses.

Finally, the model prediction error (RMSE) values of the MLR models were calculated. When the model prediction error values were compared between the two algorithms, the QA algorithm showed better or equal performance to that of XGBoost under the following six feature conditions: signal mean feature, 10 s window; signal mean feature, 60 s window; signal minimum feature, 60 s window; signal peak feature, 60 s window; signal variance feature, 10 s window; and signal variance feature, 60 s window. Tables 8 and 9 show the RMSE values for each model and time window.

Based on the results of the three-step validation process, we selected the cognitive tasks that were most highly associated with CSAT scores based on the ranked features under the following conditions: 1) signal mean, signal minimum, signal peak, and signal variance features and 2) 60-second windows. Table 10 lists the ranked features for the selected conditions. Overall, the VFT, recognition, and CBT tasks were the most frequently selected of the eight cognitive tasks.

## Additional analysis: Group comparison

For an additional analysis, we divided the participants into two groups based on their CSAT scores, and subsequently compared the results of the selected features from each group. Overall, the group with higher CSAT scores performed the ToL and N-back tasks more frequently in the top 10 orders than the group with lower scores. Tables A and B in the Supplementary Materials show the top 10 results of cognitive task variables for the 10 s and 60 s windows, comparing the results from both groups.

Using the selected feature list of the XGBoost algorithm, the results were compared between two groups with different CSAT score populations. Overall, the Stroop, CBT, and ToL task variables were highly associated with the dependent variable in the group with higher CSAT scores than in the group with lower scores. Tables C and D in the Supplementary Materials show the top 10 results of cognitive task variables for the 10 s and 60 s windows, comparing the results from both groups.

Applying the same three-step validation process with group comparison data, the following feature variables (i.e., signal slope, 60 s window; signal minimum, 60 s window) in Table 11 showed the highest relatedness with average CSAT scores in each group.

# Discussion

The feature selection results from the quantum annealing and classical machine learning models were compared to investigate the potential use of QA-based feature selection algorithms in neuroimaging data analysis. An fNIRS dataset collected during the eight cognitive tasks was used to extract 12 features for each of the two time-window conditions. Subsequently, the three-step validation process using multiple linear regression models was used for the comparison. The final cognitive tasks were selected based on their association with CSAT scores. Furthermore, we conducted a group comparison analysis by dividing the original dataset into two groups based on the average CSAT scores.

In the first validation process, the correlation coefficient values between the feature variables and CSAT scores were calculated. Previous researchers[46,47] have employed correlation coefficient values to investigate the relatedness between each independent and dependent variable. As expected, there was a trend toward a higher correlation between the highly ranked feature variables and average CSAT scores. The results show that the rank order of the selected features indicated the relative importance of the feature variables. Here, we should note that we did not consider the statistical significance of each correlation, but rather tried to obtain an overall view of the increasing/decreasing trend of the correlation coefficient values for each feature selection method and in a given time window.

In the second step, comparing the model fitness (adjusted $R^2$) values of each model, different time window conditions were taken into account. In the 60-second window condition (compared to 10-second), the QA model showed a higher scale of adjusted $R^2$ values, which made it more comparable to the CML model. Previous studies using feature selection methods through machine learning have shown that time windows with longer temporal ranges provide a better performance.[48,49] Thus, in our experimental results, QA algorithms could offer more reliable results in the longer window condition (i.e., 60-second than 10-second window), and this indicates that the feature selection results of QA algorithms represent a similar trend to that of the CML algorithms.

Finally, the model prediction errors (RMSE) of the QA and CML models were compared. The RMSE values of the QA algorithms were lower than or similar to those of the CML algorithms under the four feature variable conditions (i.e., signal mean, signal minimum, signal peak, and signal variance). For the signal mean feature variable, the RMSE values were similar across the two models. Signal mean features have been shown to be important in previous studies analyzing fNIRS data using CML models.[50,51] Furthermore, under the feature variable conditions of the signal minimum, signal peak, and signal variance, the RMSE values of the QA model were lower than those of the CML model. These three features are widely used in the feature selection process of CML models.[38,52,53] Thus, the results indicate that the selected feature sets of the QA algorithms are comparable to those of the CML model (i.e., XGBoost regressor) algorithms under the four feature variable conditions.

Finally, we selected cognitive tasks based on the feature selection results of four feature variable conditions in a 60-second time window condition. The VFT, recognition, and CBT tasks were most closely associated with academic achievement (i.e., mean CSAT scores). In agreement with our results, previous studies have found a strong link between verbal fluency and academic achievement.[54,55] Basic recognition memory capacity is a predictor of academic achievement.[56] In one study, Siquara and colleagues[57] showed that working memory performance measured using the Corsi block tapping task also successfully predicted academic achievement. In summary, three cognitive tasks (i.e., VFT, recognition, and CBT) that are highly linked to academic achievement were selected by the QA algorithms.

In addition, we conducted a group comparison analysis to compare the results between two separate datasets: a group with higher CSAT scores and a group with lower scores. The same three-step validation process was used in this study. In terms of the correlation coefficient values, only a few feature variables showed congruent trends. For the signal mean, signal root mean square, signal skewness, signal kurtosis, signal slope, signal minimum, and number of signal peaks, the correlation coefficient values were higher between the more highly ranked feature variables and the dependent variable. Thus, seven feature variables were considered when analyzing the grouped data.

Comparing model fitness (adjusted $R^2$) values, we found that the adjusted $R^2$ values of QA algorithms were higher than those of CML algorithms in the following feature variable conditions: signal mean (60-second window), signal minimum (60-second window), and signal slope (60-second window). Considering the model prediction error (RMSE) values together, the signal minimum (60-second window) and signal slope (60-second window) feature conditions were selected. Here, again, the results show that longer window range leads to better performance.[58] Signal minima and slope features have been used in neural hemodynamic data analysis using machine-learning models.[59-61]

For the final cognitive tasks, the ToL and N-back tasks were more frequently selected by the group with higher CSAT scores than by their lower counterparts. This indicates that these tasks were more highly associated with the academic achievement of the group with higher CSAT scores than those with lower scores. Both tasks are known predictors of certain kinds of academic and cognitive ability. The Tower of London task, widely known as a measure of executive planning ability, is associated with academic achievement, particularly in mathematics.[62] Luo and Zhou[63] predicted children's academic achievement levels using EEG biomarkers during N-back tasks.

This study has several implications. First, we applied QA-based algorithms to the feature selection process of neural hemodynamics (i.e., fNIRS) data. The results indicate that QA-based algorithms are comparable to CML algorithms and are feasible for multidimensional neural hemodynamic data analysis. Second, specific cognitive tasks (i.e., recognition, verbal fluency, Tower of London, and N-back) were selected based on the relative importance of feature variables and showed their associations with students' academic achievement levels (i.e., averaged CSAT scores). While the behavioral performance of cognitive tasks reflects segments of students' cognitive abilities, brain activation pattern analysis can offer an additional perspective. Furthermore, we conducted a group comparison analysis between two groups with distinct academic achievement levels, and the feature selection results suggested a clear difference. Thus, this study provides a basis for future studies investigating the potential predictors of academic achievement regarding neural activation during cognitive task phases.

This study had several limitations. First, the quantum annealing based quantum hardwares can include noises in the NISQ ('noisy intermediate-scale quantum') era.[64] Nevertheless, previous studies have reported the utility of QA-based algorithms in feature-selection tasks using neuroimaging data. Second, our fNIRS dataset was comprised of college students from South Korea. Such findings can only be generalized when samples are collected from various sources. Third, while investigating the cognitive tasks most related to academic achievement levels, we could not fully consider the individual characteristics of each cognitive task. For example, the duration of each task block was not controlled. In future studies, we plan to focus on this issue. Fourth, we used only CML models and not deep learning models to validate the results of the QA-based algorithms. Deep learning algorithms are increasingly employed for multivariate analysis; however, they require additional steps and effort to obtain feature importance values from the models. Therefore, we limited our validation to the CML models. Further studies should directly compare the feature selection results from both deep learning and QA-based algorithms.

## Conclusion

In this study, we investigated how quantum annealing algorithms can be applied in the feature selection process using the fNIRS dataset, which has mostly been conducted with classical machine learning models utilizing D-Wave quantum hardware in prior studies. Three cognitive tasks (VFT, recognition, and CBT) were found to have a relatively high correlation with students' scholastic abilities (CSAT scores). Our study demonstrates the potential of QA-based algorithms in the feature selection process by analyzing multidimensional brain hemodynamic data. Future studies should employ these results to investigate predictors of academic achievement or, more broadly, cognitive ability.

## Acknowledgements:

We would like to thank Editage (www.editage.co.kr) for editing and reviewing this manuscript for English language.

## Funding:

No.

## Conflicts of Interest:

No.

## Ethical declarations:

This study was approved by the Institutional Review Board of Yonsei University (7001988–202104-HR-659-06), and all participants provided informed consent to participate.

**Figure Legends**

Figure 1. The overview of the study procedure.

Figure 2. Schema showing the optode positions in the NIRSIT Lite device.

Figure 3. Overview of the 10-second time window slicing process for feature extraction.n

Figure 4. Correlation coefficient trend of selected features and average CSAT scores (QA-based, total data, 10-second window).

Figure 5. Correlation coefficient trend of selected features and average CSAT scores (QA-based, total data, 60-second window).

Figure 6. Correlation coefficient trend of selected features and average CSAT scores (CML, total data, 10-second window).

Figure 7. Correlation coefficient trend of selected features and average CSAT scores (CML, total data, 60-second window).

## Tables

Table 1. RMSE values of the CML models

| Feature | Time window | Root Mean Square Error (RMSE) | | |
|---|---|---|---|---|
| | | Support vector machine | Decision tree regressor | XGBoost regressor |
| Signal Mean | 10 sec | 1.458 | 1.693 | 1.001 |
| | 60 sec | 1.458 | 1.723 | 1.205 |
| Signal Median | 10 sec | 1.458 | 1.678 | 1.114 |
| | 60 sec | 1.458 | 1.751 | 1.274 |
| Signal Variance | 10 sec | 1.458 | 1.654 | 1.140 |
| | 60 sec | 1.458 | 1.664 | 1.260 |
| Signal Standard Deviation | 10 sec | 1.458 | 1.637 | 1.114 |
| | 60 sec | 1.458 | 1.756 | 1.274 |
| RMS | 10 sec | 1.458 | 1.820 | 1.196 |
| | 60 sec | 1.458 | 1.794 | 1.337 |
| Signal Peak | 10 sec | 1.458 | 1.628 | 1.008 |
| | 60 sec | 1.458 | 1.842 | 1.243 |
| Signal Minimum | 10 sec | 1.458 | 1.591 | 1.013 |
| | 60 sec | 1.458 | 1.839 | 1.265 |
| Signal Slope | 10 sec | 1.458 | 1.776 | 1.181 |
| | 60 sec | 1.458 | 1.962 | 1.407 |
| Number of Signal Peaks | 10 sec | 1.472 | 1.805 | 1.152 |
| | 60 sec | 1.408 | 1.873 | 1.335 |
| Sum of Signal Peaks | 10 sec | 2.418 | 1.756 | 1.131 |
| | 60 sec | 1.814 | 1.888 | 1.282 |
| Signal Skewness | 10 sec | 1.588 | 1.856 | 1.291 |
| | 60 sec | 1.465 | 2.066 | 1.508 |
| Signal Kurtosis | 10 sec | 1.472 | 1.910 | 1.332 |
| | 60 sec | 1.442 | 1.894 | 1.423 |
| Averaged RMSE | 10 sec | 1.551 | 1.734 | 1.139 |
| | 60 sec | 1.483 | 1.838 | 1.318 |

Table 2. Top 10 features in the ranked feature lists (QA-based, total data, 10 sec window)

| Rank | Feature (extracted by QA algorithm with 10-second windows) | | | | | |
|---|---|---|---|---|---|---|
| | Mean (# of selection) | Median (# of selection) | Variance (# of selection) | Standard Deviation (# of selection) | RMS (# of selection) | Peak (# of selection) |
| 1st | N-back_12 (168) | ToL_16 (168) | CBT_2 (168) | ToL_16 (168) | CBT_8 (167) | N-back_11 (168) |
| 2nd | VFT_15 (167) | CBT_2 (167) | ToL_16 (167) | CBT_2 (167) | resting_15 (166) | Stroop_12 (167) |
| 3rd | recog_10 (166) | VFT_5 (166) | VFT_5 (165) | VFT_5 (166) | VFT_9 (166) | CBT_8 (165) |
| 4th | CBT_9 (165) | recog_9 (165) | CBT_19 (164) | recog_9 (165) | VFT_15 (165) | VFT_9 (164) |
| 5th | resting_15 (163) | recog_15 (164) | VFT_18 (164) | recog_15 (164) | VFT_4 (164) | Stroop_7 (161) |
| 6th | CBT_12 (161) | VFT_18 (161) | recog_9 (163) | recog_2 (160) | recog_9 (162) | VFT_15 (160) |
| 7th | emotion_0 (161) | VFT_15 (160) | VFT_20 (163) | VFT_18 (160) | emotion_0 (161) | emotion_0 (160) |
| 8th | recog_9 (160) | VFT_6 (160) | VFT_15 (162) | resting_1 (159) | CBT_2 (160) | ToL_9 (159) |
| 9th | CBT_8 (159) | recog_2 (159) | recog_15 (159) | VFT_15 (158) | CBT_9 (160) | VFT_12 (159) |
| 10th | CBT_13 (158) | recog_16 (157) | recog_2 (159) | VFT_6 (157) | resting_11 (159) | N-back_12 (159) |

| | Minimum (# of selection) | Slope (# of selection) | Number of Peaks (# of selection) | Sum of Peaks (# of selection) | Skewness (# of selection) | Kurtosis (# of selection) |
|---|---|---|---|---|---|---|
| 1st | VFT_17 (167) | Stroop_7 (168) | recog_4 (168) | emotion_12 (167) | Stroop_14 (166) | VFT_1 (168) |
| 2nd | VFT_4 (167) | CBT_2 (167) | Stroop_20 (167) | recog_20 (166) | recog_11 (164) | Stroop_12 (167) |
| 3rd | VFT_15 (166) | CBT_7 (166) | CBT_5 (166) | recog_3 (165) | Stroop_10 (164) | CBT_9 (166) |
| 4th | VFT_9 (163) | N-back_5 (164) | resting_4 (163) | recog_18 (162) | emotion_14 (162) | N-back_0 (164) |
| 5th | resting_15 (161) | CBT_6 (164) | Stroop_9 (163) | Stroop_12 (162) | emotion_18 (162) | CBT_0 (161) |
| 6th | CBT_1 (161) | recog_2 (163) | resting_7 (161) | emotion_9 (161) | VFT_1 (162) | N-back_9 (160) |
| 7th | CBT_8 (160) | recog_9 (161) | ToL_20 (161) | CBT_5 (161) | Stroop_12 (162) | VFT_8 (160) |
| 8th | recog_9 (159) | recog_8 (160) | emotion_0 (161) | emotion_8 (161) | N-back_12 (161) | emotion_16 (159) |
| 9th | Stroop_7 (159) | recog_11 (160) | Stroop_8 (159) | CBT_11 (160) | CBT_2 (161) | VFT_15 (159) |
| 10th | VFT_5<br>(158) | N-back_10<br>(158) | VFT_19<br>(158) | ToL_10<br>(160) | Stroop_15<br>(159) | N-back_16<br>(158) |

*Abbreviations: CBT (Corsi-block tapping task); recog (recognition task); ToL (Tower of London task); VFT (verbal fluency task).*

Table 3. Top 10 features in the ranked feature lists (QA-based, total data, 60 sec window)

| Rank | Feature (extracted by QA algorithm with 60-second windows) | | | | | |
|---|---|---|---|---|---|---|
| | Mean (# of selection) | Median (# of selection) | Variance (# of selection) | Standard Deviation (# of selection) | RMS (# of selection) | Peak (# of selection) |
| 1st | VFT_1 (24) | recog_2 (24) | Stroop_0 (24) | recog_2 (24) | recog_2 (24) | VFT_2 (23) |
| 2nd | N-back_0 (22) | VFT_2 (23) | VFT_2 (23) | VFT_2 (23) | VFT_2 (23) | Stroop_0 (23) |
| 3rd | VFT_2 (22) | N-back_1 (22) | CBT_0 (22) | N-back_1 (22) | CBT_0 (22) | resting_2 (22) |
| 4th | N-back_2 (22) | CBT_0 (21) | N-back_1 (21) | CBT_0 (21) | resting_2 (21) | recog_2 (21) |
| 5th | recog_2 (20) | Stroop_0 (20) | recog_2 (19) | Stroop_0 (20) | VFT_1 (20) | recog_1 (21) |
| 6th | CBT_2 (19) | VFT_1 (19) | resting_2 (19) | VFT_1 (19) | VFT_0 (19) | VFT_1 (19) |
| 7th | resting_2 (18) | CBT_1 (18) | VFT_1 (19) | CBT_1 (18) | Stroop_0 (18) | CBT_0 (18) |
| 8th | CBT_1 (17) | VFT_0 (17) | emotion_1 (17) | VFT_0 (17) | N-back_1 (17) | N-back_1 (17) |
| 9th | Stroop_2 (16) | resting_2 (16) | VFT_0 (16) | recog_1 (16) | resting_1 (16) | CBT_1 (17) |
| 10th | VFT_0 (15) | recog_1 (16) | recog_0 (15) | resting_2 (16) | recog_1 (16) | resting_1 (16) |
| | Minimum (# of selection) | Slope (# of selection) | Number of Peaks (# of selection) | Sum of Peaks (# of selection) | Skewness (# of selection) | Kurtosis (# of selection) |
| 1st | recog_2 (24) | N-back_2 (24) | CBT_0 (23) | emotion_2 (23) | Stroop_0 (23) | resting_2 (23) |
| 2nd | VFT_1 (23) | VFT_1 (23) | emotion_1 (23) | CBT_2 (22) | N-back_1 (23) | CBT_0 (23) |
| 3rd | CBT_0 (22) | CBT_0 (22) | emotion_2 (23) | emotion_1 (22) | CBT_0 (22) | ToL_0 (22) |
| 4th | VFT_2 (21) | VFT_2 (21) | Stroop_0 (20) | resting_2 (21) | resting_2 (21) | Stroop_0 (21) |
| 5th | resting_2 (20) | ToL_2 (20) | VFT_0 (19) | Stroop_0 (19) | VFT_0 (20) | CBT_1 (20) |
| 6th | recog_1 (19) | N-back_1 (18) | resting_0 (18) | N-back_0 (19) | resting_0 (19) | resting_1 (20) |
| 7th | resting_1 (18) | recog_0 (18) | VFT_1 (18) | ToL_1 (18) | ToL_0 (18) | N-back_2 (18) |
| 8th | N-back_1 | recog_1 | resting_1 | CBT_0 | Stroop_2 | resting_0 |
| | (17) | (18) | (17) | (17) | (18) | (17) |
| 9th | CBT_1<br>(17) | Stroop_1<br>(16) | emotion_0<br>(17) | VFT_1<br>(17) | CBT_1<br>(17) | N-back_1<br>(17) |
| 10th | recog_0<br>(15) | VFT_0<br>(16) | resting_2<br>(17) | Stroop_2<br>(16) | resting_1<br>(16) | VFT_2<br>(16) |

*Abbreviations: CBT (Corsi-block tapping task); recog (recognition task); ToL (Tower of London task); VFT (verbal fluency task).*

Table 4. Top 10 features in the ranked feature lists (CML, total data, 10 sec window)

| Rank | Feature (extracted by CML algorithm with 10-second windows) | | | | | |
|---|---|---|---|---|---|---|
| | Mean (F-score) | Median (F-score) | Variance (F-score) | Standard Deviation (F-score) | RMS (F-score) | Peak (F-score) |
| 1st | VFT_7 (5.462) | N-back_20 (13.607) | emotion_15 (9.008) | N-back_20 (13.607) | N-back_20 (9.721) | resting_18 (6.156) |
| 2nd | recog_9 (5.070) | VFT_10 (6.748) | N-back_20 (8.641) | VFT_10 (6.748) | recog_17 (7.117) | N-back_2 (5.779) |
| 3rd | ToL_11 (4.947) | resting_1 (6.548) | resting_1 (7.331) | resting_1 (6.548) | recog_9 (7.002) | emotion_3 (5.764) |
| 4th | recog_18 (4.840) | ToL_11 (6.095) | ToL_11 (7.259) | ToL_11 (6.095) | ToL_11 (5.607) | N-back_16 (5.011) |
| 5th | emotion_10 (4.826) | VFT_15 (6.056) | Stroop_12 (6.122) | VFT_15 (6.056) | VFT_13 (5.518) | VFT_13 (4.947) |
| 6th | emotion_4 (4.714) | Stroop_12 (5.254) | VFT_15 (5.918) | Stroop_12 (5.254) | emotion_18 (5.350) | resting_15 (4.138) |
| 7th | emotion_2 (4.327) | VFT_8 (5.220) | emotion_5 (5.002) | VFT_8 (5.220) | VFT_6 (5.065) | VFT_7 (4.068) |
| 8th | Stroop_9 (3.838) | resting_18 (4.756) | VFT_12 (4.567) | resting_18 (4.756) | emotion_16 (4.700) | CBT_2 (4.001) |
| 9th | VFT_6 (3.522) | recog_13 (4.533) | Stroop_3 (4.529) | recog_13 (4.533) | VFT_8 (4.506) | recog_16 (3.786) |
| 10th | recog_10 (3.351) | emotion_15 (4.506) | recog_13 (4.016) | emotion_15 (4.506) | recog_6 (4.421) | N-back_10 (3.541) |
| | Minimum (F-score) | Slope (F-score) | Number of Peaks (F-score) | Sum of Peaks (F-score) | Skewness (F-score) | Kurtosis (F-score) |
| 1st | N-back_20 (6.675) | recog_13 (9.891) | VFT_5 (7.428) | N-back_8 (9.893) | recog_5 (7.800) | VFT_0 (7.849) |
| 2nd | N-back_9 (5.287) | N-back_20 (8.700) | CBT_11 (5.998) | recog_18 (7.624) | VFT_4 (5.814) | resting_1 (6.993) |
| 3rd | emotion_19 (5.176) | ToL_11 (6.758) | recog_10 (5.759) | N-back_2 (7.464) | VFT_7 (5.625) | VFT_12 (6.547) |
| 4th | VFT_8 (4.660) | recog_7 (6.447) | N-back_2 (5.438) | N-back_11 (7.035) | N-back_20 (5.624) | emotion_17 (6.534) |
| 5th | Stroop_10 (4.639) | emotion_15 (6.390) | Stroop_12 (5.256) | VFT_5 (6.800) | recog_6 (5.623) | recog_16 (6.082) |
| 6th | ToL_11 (4.428) | Stroop_19 (5.464) | CBT_20 (4.746) | resting_3 (6.644) | CBT_13 (5.389) | recog_12 (5.872) |
| 7th | VFT_13 (4.155) | resting_15 (5.168) | Stroop_4 (4.685) | emotion_16 (6.216) | recog_2 (5.301) | emotion_18 (5.810) |
| 8th | VFT_16 | resting_13 | recog_15 | recog_20 | recog_10 | Stroop_5 |
| | (3.963) | (5.086) | (4.499) | (6.153) | (5.245) | (5.074) |
| 9th | VFT_11<br>(3.894) | Stroop_14<br>(4.747) | VFT_1<br>(4.322) | recog_7<br>(5.735) | N-back_9<br>(5.046) | recog_20<br>(5.024) |
| 10th | emotion_10<br>(3.828) | Stroop_12<br>(4.620) | N-back_7<br>(4.306) | emotion_0<br>(5.489) | emotion_17<br>(4.988) | Stroop_12<br>(4.951) |

*Abbreviations: CBT (Corsi-block tapping task); recog (recognition task); ToL (Tower of London task); VFT (verbal fluency task).*

Table 5. Top 10 features in the ranked feature lists (CML, total data, 60 sec window)

| Rank | Feature (extracted by CML algorithm with 60-second windows) | | | | | |
|---|---|---|---|---|---|---|
| | Mean (F-score) | Median (F-score) | Variance (F-score) | Standard Deviation (F-score) | RMS (F-score) | Peak (F-score) |
| 1st | VFT_1 (2.743) | recog_1 (3.174) | recog_1 (2.998) | recog_1 (3.174) | recog_2 (3.207) | N-back_0 (2.597) |
| 2nd | emotion_0 (2.656) | recog_2 (2.915) | VFT_2 (2.736) | recog_2 (2.915) | recog_1 (3.002) | Stroop_1 (2.506) |
| 3rd | recog_2 (2.323) | VFT_2 (2.318) | N-back_1 (2.546) | VFT_2 (2.318) | emotion_0 (2.531) | VFT_2 (2.473) |
| 4th | Stroop_2 (2.315) | ToL_2 (2.239) | recog_2 (2.485) | ToL_2 (2.239) | N-back_2 (2.340) | emotion_1 (2.448) |
| 5th | Stroop_1 (2.040) | N-back_1 (2.110) | emotion_1 (2.329) | N-back_1 (2.110) | VFT_2 (2.203) | CBT_1 (2.427) |
| 6th | recog_1 (1.997) | N-back_2 (2.101) | CBT_1 (2.170) | N-back_2 (2.101) | Stroop_2 (1.975) | Stroop_2 (2.424) |
| 7th | VFT_2 (1.941) | VFT_1 (1.992) | Stroop_2 (1.950) | VFT_1 (1.992) | VFT_1 (1.974) | emotion_2 (2.252) |
| 8th | resting_2 (1.925) | emotion_0 (1.930) | emotion_0 (1.799) | emotion_0 (1.930) | N-back_1 (1.951) | VFT_1 (2.184) |
| 9th | recog_0 (1.879) | CBT_1 (1.921) | N-back_2 (1.774) | CBT_1 (1.921) | recog_0 (1.938) | VFT_0 (2.150) |
| 10th | Stroop_0 (1.871) | emotion_1 (1.876) | recog_0 (1.758) | emotion_1 (1.876) | emotion_2 (1.877) | recog_0 (2.111) |
| | Minimum (F-score) | Slope (F-score) | Number of Peaks (F-score) | Sum of Peaks (F-score) | Skewness (F-score) | Kurtosis (F-score) |
| 1st | VFT_2 (2.840) | N-back_2 (2.730) | VFT_1 (2.470) | emotion_2 (3.033) | N-back_2 (3.486) | VFT_0 (3.551) |
| 2nd | VFT_1 (2.830) | emotion_2 (2.696) | emotion_1 (2.038) | VFT_0 (2.874) | N-back_1 (3.049) | resting_1 (3.443) |
| 3rd | emotion_2 (2.710) | recog_2 (2.646) | CBT_0 (1.867) | Stroop_0 (2.359) | resting_2 (2.569) | VFT_12 (3.181) |
| 4th | recog_0 (2.312) | ToL_2 (2.551) | VFT_0 (1.854) | emotion_1 (2.243) | Stroop_0 (2.559) | emotion_17 (2.805) |
| 5th | emotion_1 (2.300) | Stroop_1 (2.403) | emotion_2 (1.845) | emotion_0 (2.192) | resting_0 (2.457) | recog_16 (2.799) |
| 6th | recog_1 (2.198) | emotion_1 (2.386) | resting_1 (1.836) | VFT_2 (2.180) | recog_2 (2.415) | recog_12 (2.585) |
| 7th | recog_2 (2.194) | recog_1 (2.322) | Stroop_0 (1.818) | VFT_1 (2.026) | VFT_1 (2.398) | emotion_18 (2.565) |
| 8th | emotion_0 | VFT_2 | recog_1 | CBT_2 | emotion_0 | Stroop_5 |
| | (2.188) | (2.219) | (1.800) | (1.987) | (2.345) | (2.537) |
| 9th | CBT_2<br>(2.152) | recog_0<br>(2.205) | VFT_2<br>(1.796) | resting_1<br>(1.973) | VFT_0<br>(2.326) | recog_20<br>(2.449) |
| 10th | N-back_0<br>(2.008) | VFT_0<br>(2.185) | resting_2<br>(1.700) | N-back_2<br>(1.934) | CBT_0<br>(2.269) | Stroop_12<br>(2.353) |

*Abbreviations: CBT (Corsi-block tapping task); recog (recognition task); ToL (Tower of London task); VFT (verbal fluency task).*

Table 6. Model fitness (adjusted $R^2$) values of the QA and XGBoost models with 10-second windows

Model fitness values of each feature (extracted with 10-second windows)

| Rank | Mean | | Median | | Variance | | Standard Deviation | |
|---|---|---|---|---|---|---|---|---|
| | QA | CML | QA | CML | QA | CML | QA | CML |
| top 10 | 0.014 (0.013) | 0.007 (0.006) | 0.05 (0.05) | 0.082 (0.082) | 0.015 (0.014) | 0.018 (0.018) | 0.055 (0.054) | 0.082 (0.082) |
| top 20 | 0.027 (0.027) | 0.018 (0.016) | 0.066 (0.065) | 0.091 (0.09) | 0.024 (0.023) | 0.032 (0.03) | 0.064 (0.063) | 0.091 (0.09) |
| top 30 | 0.036 (0.035) | 0.024 (0.021) | 0.080 (0.079) | 0.109 (0.107) | 0.031 (0.029) | 0.051 (0.049) | 0.08 (0.079) | 0.109 (0.107) |
| top 40 | 0.049 (0.047) | 0.033 (0.032) | 0.100 (0.098) | 0.128 (0.126) | 0.045 (0.043) | 0.055 (0.054) | 0.100 (0.098) | 0.128 (0.126) |
| top 50 | 0.063 (0.058) | 0.044 (0.041) | 0.124 (0.122) | 0.153 (0.151) | 0.048 (0.046) | 0.065 (0.061) | 0.12 (0.118) | 0.153 (0.151) |

| Rank | RMS | | Peak | | Minimum | | Slope | |
|---|---|---|---|---|---|---|---|---|
| | QA | CML | QA | CML | QA | CML | QA | CML |
| top 10 | 0.043 (0.043) | 0.066 (0.066) | 0.017 (0.017) | 0.025 (0.025) | 0.009 (0.009) | 0.045 (0.045) | 0.02 (0.02) | 0.095 (0.095) |
| top 20 | 0.054 (0.053) | 0.086 (0.085) | 0.026 (0.025) | 0.033 (0.031) | 0.023 (0.022) | 0.057 (0.056) | 0.049 (0.048) | 0.126 (0.125) |
| top 30 | 0.074 (0.073) | 0.11 (0.11) | 0.051 (0.049) | 0.054 (0.053) | 0.033 (0.03) | 0.065 (0.063) | 0.054 (0.051) | 0.14 (0.139) |
| top 40 | 0.088 (0.085) | 0.132 (0.131) | 0.067 (0.065) | 0.079 (0.076) | 0.045 (0.041) | 0.073 (0.07) | 0.068 (0.064) | 0.165 (0.164) |
| top 50 | 0.099 (0.096) | 0.147 (0.145) | 0.077 (0.074) | 0.09 (0.087) | 0.074 (0.071) | 0.086 (0.084) | 0.118 (0.114) | 0.176 (0.175) |

| Rank | Number of Peaks | | Sum of Peaks | | Skewness | | Kurtosis | |
|---|---|---|---|---|---|---|---|---|
| | QA | CML | QA | CML | QA | CML | QA | CML |
| top 10 | 0.004 (0.004) | 0.037 (0.036) | 0.013 (0.012) | 0.032 (0.031) | 0.004 (0.002) | 0.021 (0.021) | 0.006 (0.004) | 0.053 (0.053) |
| top 20 | 0.03 (0.03) | 0.053 (0.053) | 0.016 (0.014) | 0.054 (0.053) | 0.013 (0.011) | 0.033 (0.033) | 0.027 (0.025) | 0.068 (0.069) |
| top 30 | 0.042 (0.041) | 0.066 (0.066) | 0.027 (0.024) | 0.074 (0.072) | 0.015 (0.012) | 0.047 (0.046) | 0.046 (0.044) | 0.09 (0.089) |
| top 40 | 0.053<br>(0.051) | 0.076<br>(0.074) | 0.037<br>(0.033) | 0.083<br>(0.08) | 0.022<br>(0.017) | 0.05<br>(0.047) | 0.063<br>(0.061) | 0.119<br>(0.118) |
| top 50 | 0.06<br>(0.058) | 0.08<br>(0.077) | 0.044<br>(0.04) | 0.095<br>(0.092) | 0.026<br>(0.02) | 0.056<br>(0.052) | 0.079<br>(0.077) | 0.133<br>(0.131) |

Table 7. Model fitness (adjusted $R^2$) values of the QA and XGBoost models with 60-second windows

| Rank | Model fitness values of each feature (extracted with 60-second windows) | | | | | | | |
|---|---|---|---|---|---|---|---|---|
| | Mean | | Median | | Variance | | Standard Deviation | |
| | QA | CML | QA | CML | QA | CML | QA | CML |
| top 10 | 0.016 (0.015) | 0.016 (0.014) | 0.045 (0.045) | 0.049 (0.048) | 0.02 (0.019) | 0.013 (0.012) | 0.045 (0.045) | 0.049 (0.048) |
| top 20 | 0.019 (0.016) | 0.019 (0.016) | 0.073 (0.073) | 0.073 (0.073) | 0.033 (0.031) | 0.025 (0.024) | 0.073 (0.073) | 0.073 (0.073) |
| | RMS | | Peak | | Minimum | | Slope | |
| | QA | CML | QA | CML | QA | CML | QA | CML |
| top 10 | 0.042 (0.041) | 0.041 (0.041) | 0.021 (0.02) | 0.011 (0.011) | 0.028 (0.027) | 0.015 (0.014) | 0.041 (0.041) | 0.04 (0.04) |
| top 20 | 0.067 (0.067) | 0.067 (0.066) | 0.041 (0.04) | 0.032 (0.031) | 0.051 (0.051) | 0.045 (0.044) | 0.054 (0.054) | 0.05 (0.049) |
| | Number of Peaks | | Sum of Peaks | | Skewness | | Kurtosis | |
| | QA | CML | QA | CML | QA | CML | QA | CML |
| top 10 | 0.033 (0.033) | 0.034 (0.034) | 0.014 (0.014) | 0.023 (0.022) | 0.009 (0.009) | 0.009 (0.009) | 0.029 (0.028) | 0.039 (0.04) |
| top 20 | 0.053 (0.053) | 0.056 (0.057) | 0.026 (0.025) | 0.037 (0.037) | 0.012 (0.01) | 0.012 (0.01) | 0.059 (0.058) | 0.059 (0.058) |

Table 8. Model prediction error (RMSE) values of the QA and XGBoost models with 10-second windows

| Rank | Model prediction error values of each feature (extracted with 10-second windows) | | | | | | | |
|---|---|---|---|---|---|---|---|---|
| | Mean | | Median | | Variance | | Standard Deviation | |
| | QA | CML | QA | CML | QA | CML | QA | CML |
| top 10 | **1.453** | 1.467 | 1.434 | 1.410 | **1.465** | 1.493 | 1.440 | 1.410 |
| top 20 | **1.448** | 1.469 | 1.433 | 1.423 | **1.470** | 1.507 | 1.436 | 1.423 |
| top 30 | **1.454** | 1.473 | 1.422 | 1.410 | **1.480** | 1.526 | 1.422 | 1.410 |
| top 40 | **1.446** | 1.474 | 1.421 | 1.398 | **1.478** | 1.617 | 1.421 | 1.398 |
| top 50 | **1.454** | 1.467 | 1.422 | 1.390 | **1.493** | 1.570 | 1.418 | 1.390 |
| | RMS | | Peak | | Minimum | | Slope | |
| | QA | CML | QA | CML | QA | CML | QA | CML |
| top 10 | 1.442 | 1.418 | 1.460 | 1.447 | 1.473 | 1.464 | 1.463 | 1.408 |
| top 20 | 1.433 | 1.403 | 1.459 | 1.440 | 1.472 | 1.460 | 1.443 | 1.405 |
| top 30 | 1.426 | 1.401 | 1.432 | 1.419 | 1.465 | 1.469 | 1.446 | 1.408 |
| top 40 | 1.435 | 1.382 | 1.432 | 1.425 | 1.449 | 1.462 | 1.437 | 1.400 |
| top 50 | 1.421 | 1.391 | 1.445 | 1.424 | 1.447 | 1.465 | 1.408 | 1.388 |
| | Number of Peaks | | Sum of Peaks | | Skewness | | Kurtosis | |
| | QA | CML | QA | CML | QA | CML | QA | CML |
| top 10 | 1.475 | 1.443 | 1.478 | 1.455 | 1.470 | 1.461 | 1.467 | 1.447 |
| top 20 | 1.456 | 1.424 | 1.481 | 1.442 | 1.473 | 1.453 | 1.467 | 1.449 |
| top 30 | 1.448 | 1.416 | 1.469 | 1.425 | 1.472 | 1.444 | 1.455 | 1.439 |
| top 40 | 1.445 | 1.421 | 1.467 | 1.428 | 1.474 | 1.451 | 1.445 | 1.419 |
| top 50 | 1.447 | 1.423 | 1.462 | 1.432 | 1.473 | 1.451 | 1.435 | 1.414 |

Table 9. Model prediction error (RMSE) values of the QA and XGBoost models with 60-second windows

| Rank | Model prediction error values of each feature (extracted with 60-second windows) | | | | | | | |
|---|---|---|---|---|---|---|---|---|
| | Mean | | Median | | Variance | | Standard Deviation | |
| | QA | CML | QA | CML | QA | CML | QA | CML |
| top 10 | **1.464** | 1.463 | 1.435 | 1.431 | **1.450** | 1.461 | 1.435 | 1.431 |
| top 20 | **1.463** | 1.463 | 1.415 | 1.413 | **1.439** | 1.450 | 1.415 | 1.413 |
| | RMS | | Peak | | Minimum | | Slope | |
| | QA | CML | QA | CML | QA | CML | QA | CML |
| top 10 | 1.441 | 1.442 | **1.461** | 1.468 | **1.447** | 1.464 | 1.451 | 1.452 |
| top 20 | 1.417 | 1.417 | **1.457** | 1.473 | **1.428** | 1.434 | 1.447 | 1.444 |
| | Number of Peaks | | Sum of Peaks | | Skewness | | Kurtosis | |
| | QA | CML | QA | CML | QA | CML | QA | CML |
| top 10 | 1.468 | 1.464 | 1.470 | 1.463 | 1.489 | 1.481 | 1.469 | 1.452 |
| top 20 | 1.455 | 1.459 | 1.475 | 1.463 | 1.497 | 1.493 | 1.444 | 1.443 |

Table 10. Top 10 ranked features of the selected conditions (signal mean, signal minimum, signal peak, signal variance; 60-second window)

| Rank | Feature (extracted by QA algorithm with 60-second windows) | | | |
|---|---|---|---|---|
| | Mean (# of selection) | Minimum (# of selection) | Peak (# of selection) | Variance (# of selection) |
| 1st | VFT_1 (24) | recog_2 (24) | VFT_2 (23) | Stroop_0 (24) |
| 2nd | N-back_0 (22) | VFT_1 (23) | Stroop_0 (23) | VFT_2 (23) |
| 3rd | VFT_2 (22) | CBT_0 (22) | resting_2 (22) | CBT_0 (22) |
| 4th | N-back_2 (22) | VFT_2 (21) | recog_2 (21) | N-back_1 (21) |
| 5th | recog_2 (20) | resting_2 (20) | recog_1 (21) | recog_2 (19) |
| 6th | CBT_2 (19) | recog_1 (19) | VFT_1 (19) | resting_2 (19) |
| 7th | resting_2 (18) | resting_1 (18) | CBT_0 (18) | VFT_1 (19) |
| 8th | CBT_1 (17) | N-back_1 (17) | N-back_1 (17) | emotion_1 (17) |
| 9th | Stroop_2 (16) | CBT_1 (17) | CBT_1 (17) | VFT_0 (16) |
| 10th | VFT_0 (15) | recog_0 (15) | resting_1 (16) | recog_0 (15) |

*Abbreviations: CBT (Corsi-block tapping task); recog (recognition task); ToL (Tower of London task); VFT (verbal fluency task).*

Table 11. Top 10 ranked features of selected conditions (signal slope, signal minimum; 60-second window)

| Rank | Feature (extracted by QA algorithm with 60-second windows) | | | |
|---|---|---|---|---|
| | Minimum (# of selection) | | Slope (# of selection) | |
| | g1 | g2 | g1 | g2 |
| 1st | VFT_1 (24) | CBT_2 (23) | N-back_1 (24) | N-back_2 (24) |
| 2nd | N-back_2 (23) | recog_1 (22) | N-back_0 (23) | CBT_0 (22) |
| 3rd | N-back_1 (22) | recog_2 (22) | N-back_2 (22) | recog_1 (22) |
| 4th | recog_0 (21) | resting_2 (22) | Stroop_1 (21) | VFT_0 (22) |
| 5th | recog_2 (20) | CBT_0 (21) | resting_2 (20) | N-back_1 (20) |
| 6th | VFT_2 (19) | VFT_1 (19) | CBT_0 (19) | recog_0 (19) |
| 7th | N-back_0 (18) | CBT_1 (18) | VFT_2 (18) | Stroop_1 (18) |
| 8th | recog_1 (17) | recog_0 (17) | recog_2 (17) | Stroop_0 (17) |
| 9th | emotion_2 (16) | resting_1 (16) | ToL_0 (17) | VFT_2 (16) |
| 10th | ToL_0 (15) | N-back_1 (16) | ToL_1 (16) | VFT_1 (16) |